\documentclass [12pt]{article}
\usepackage{amsmath}
\usepackage[dvips]{graphicx}
\begin{document}

\begin{center}
\Large{The Earth's figure axis determined from the polar motion data}\\
~\\
\large{Yoshio Kubo} \normalsize{(kuboy@sakura.to)\\
22 June 2012}
\end{center}

\begin{center}
\textbf{Abstract}
\end{center}

The polar motion data is analyzed to obtain accurate position of the figure axis referred to the Earth-fixed frame.
The variation of the figure axis should be the basic object to which the geophysical events are linked.
By the method of rigid dynamics, % the position and motion of the rotational axis and the figure axis in the Earth can be determined accurately,
% regarding the elasticity and anelasticity of the Earth as perturbations to the rotation of a rigid Earth.  Using the result, 
the relation between the rotational and the figure axes is derived.
The polar motion is the motion of the rotational axis on the Earth's surface and therefore
the exact position of the figure axis on the surface at each moment is obtained from the polar motion.
Since the accuracy of the recent data for the polar motion is very high, the obtained position of the figure axis is considered to keep good quality.
As an average for a long duration of time, the obtained figure axis exhibits stable annual and semi-annual variations
while it has no component with the period of the Chandler wobble.
Besides this general feature, it shows different patterns from year to year as well as small irregularities with shorter periods,
most of which are considered to be significant but not attributed to the observational errors. 
Further, a simple model is introduced for the cause of the seasonal variation of the figure axis.
The model explains both the variation of the figure axis obtained above and that of the rotational speed which is so far known.

~

\section{Introduction}

The polar motion is a phenomenon in the rotational motion of the Earth,
appearing as the motion of the point on the Earth's surface at which the rotational axis pierces the surface.  The motion is roughly circular but with changing radius.

This phenomenon occurs in the free rotation of a rigid Earth as well.
When the Earth is regarded as a rigid oblate spheroid, the locus is a complete circle.
The period of the circular motion in the rigid case is $A/(C - A)$ times the period of the  Earth's rotation,
$C$ and $A$ being the moments of inertia of the Earth around its symmetrical axis and around an axis on the equator of the symmetrical axis, respectively.
We suppose the approximation that the Earth is a spheroid throughout the present study.

For the value of $A/(C - A)$ in the actual Earth, the period is about 304 sidereal days.
The motion of the rotational axis in a rigid Earth is called the Euler motion and its existence and the behavior can be proved theoretically.

In the actual Earth, the Euler motion suffers a considerable deformation.  The motion is now not with a simple shape and with a single period
but a compound of the various elliptical motions with different periods, shapes and sizes, all varying with time.

The largest ellipse among them stems from the Euler motion in the rigid case and is called the free oscillation (Munk and McDonald, 1960).
The period of of the free oscillation is about 435 days, considerably longer than that in the rigid Earth.
The free oscillation is also called the Chandler wobble after the name of the discoverer of the period.

The prolongation of the period in the free oscillation in the actual Earth is attributed to the elasticity of the Earth.
It was first explained by Newcomb (1892).  Explanations are also seen in Munk and McDonald (1960) as well as in Kubo (1991).
As it is important to take the prolongation of the period into account in the argument in Section 2, a detailed review of it is given there.

Besides the Chandler wobble and the many other periodical components in the polar motion is also found a secular drift of its center.

All the motions other than the free oscillation are considered to be reflecting the variation of the mass distribution
in the whole Earth including the ocean and the atmosphere.
Therefore, the investigation of the polar motion has held a very important position for understanding the dynamical state of the Earth.
 
For that reason, numerous analyses have been published concerning the polar motion, especially for its periodic parts.
Among them, some studies aim only to analyze the pattern of the polar motion itself such as the variations in its amplitude and phase,
e.g., Kimura (1917), Guinot (1972), H$\ddot {\textrm{o}}$pfner (2002).
Some other studies intend to explain the variation in the polar motion by geophysical causes such as the atmosphere (Aoyama and Naito, 2001),
the ocean (Dickman, 1993; Gross, 2000), geomagnetic jerks (Gibert, 2008) and so on.

The polar motion, however, is considered to reflect the state of the Earth not so directly as the variation of the figure axis.
The change in the inner state of the Earth first brings about a variation in the figure axis, which then affects the polar motion.
Therefore it is the variation of the figure axis on the Earth's surface but not the polar motion that geophysical events should be linked to.

The present study intends to obtain the position and the motion of the figure axis relative to the Earth's surface. 
It is performed by the method of rigid dynamics,
regarding the elasticity and anelasticity of the Earth as perturbations to the rotation of a rigid Earth.

Primarily, a main theme of the rotation of the Earth in rigid dynamics is to obtain the respective motions of the rotational axis and the figure axis.
The solutions to the rotation of the rigid Earth as well as of the elastic Earth have been published numerously so far,
among which a typical example for the former is Kinoshita (1977) and one for the latter would be Kubo (1991).

It is not necessary at present, however, to solve the respective motions of both axes.
% Most important information we need in understanding the status of the Earth is the position of the figure axis at any moment.
We only have to determine the position of the figure axis at any moment on the Earth's surface
and we can attain this purpose only if the relation between the rotational and the figure axes is known
because the observed polar motion is the position of the rotational axis on the Earth's surface.
Both the figure axis and the rotational axis are definite physical entities and the relation between them can be determined without any ambiguity.

The derivation of the fundamental equations to give the relation between both axes is carried out in Section 2 based on the result in Kubo (1991), 
and some preliminary studies using the equations are presented in Sections 3 and 4.

In Sections  5 and 6, the actual polar motion of the Earth is exermined.
First in Section 5 are overviewed the feature of the polar motion published by IERS 
as well as the spectrum distribution for the components contained in it,
with a result that all the conclusions reached are not different largely from those preceding.

In Section 6, on the other hand, is presented a quite new result on the position and the motion of the figure axis which are obtained from the data of the polar motion.
 
Finally, in Section 7, a simple model of the mass transfer in the Earth is introduced to examine if the variation of the figure axis obtained above is
reasonable or not. 

\section{Equations for the polar motion}

\subsection{Definitions of the figure axis and the rotational axis in the present study}

The polar motion in the rotation of the Earth is defined as the motion of the rotational axis referred to a coordinate system set on the Earth surface
near the north pole.
We separate this motion into two parts:  One is the motion of the Earth's figure axis in the said coordinate system
and the other is the motion of the rotational axis relative to the moving figure axis.
The motion of the figure axis does not exist in the rotation of the rigid Earth but only the latter motion.

In the deformable Earth, where the variation of the figure axis appears, it is important to give exact definitions for both the figure axis and the rotational axis.
Therefore, first we consider the rigorous definition for both axes adopted in the present study.

As for the rotational axis in the present study we adopt Celestial Intermediate Pole (CIP) which is defined definitely in the resolution of the International Astronomical
Union (IAU, 2000).

CIP, which was called Celestial Ephemeris Pole (CEP) before, defines the fundamental coordinate system of date in the celestial sphere.
The precession and the nutation in the astronomical ephemeris are defined as the secular and the periodic motions of CIP, respectively.  
CIP is the axis around which the Earth actually makes the diurnal motion
but different from the rotational axis in mathematical sense by less than $0.01''$ (Seidelmann, 1982; IERS, 2004).

% On the celestial sphere the point representing the direction of the figure axis makes a diurnal
% circular motion around the point corresponding to that of CIP.  This fact implies that the Earth is actually rotating around CIP.

% As a consequence that CIP is moving due to the precession and nutation at the same time, however, CIP is not the rotational axis in the mathematical sense.
% The mathematical rotational axis is displaced about several tens milliseconds of arc from CIP, the displacement being not constant with time.
% In order to distinguish both the axes, CIP is called the averaged figure axis or simply the figure axis in some cases, especially in the theory of the precession and nutation.
% It should, however, never be confused with the figure axis defined below.

The motion of the rotational axis defined by CIP does not contain diurnal component in space nor with respect to the surface of the Earth.
Therefore the polar motion defined by this rotational axis does not have diurnal component.
The coordinates of the pole in the polar motion published from IERS are based on this axis.

Next we define the figure axis in the present study.
The figure axis of the Earth which is changing at every instant is the axis with the largest moment of inertia at this instant.
In the present study, however, we do not consider changes with short period close to the diurnal motion, for example due to the body tide and others (if they exist)
but the axis averaged as for the periodic changes, say, shorter than several days.

More importantly, however, we must take into account the fact that the actual figure axis at any instant is the axis with the largest moment of inertia
formed as the compound of the Earth's figure which the Earth would have if it were not rotating, i.e., the intrinsic figure of the Earth,
and the bulge produced by the centrifugal force due to the rotation which is formed symmetrically around the rotational axis (the rotational axis defined just above).
In the following we call this axis for the compounded figure of the Earth the effective figure axis.

The effective axis displaces from the intrinsic axis by an amount explained later.
This displacement does not occur with a short period but a comparably long period, synchronous with the period of the Chandler wobble.
However, we do not let this displacement of the effective figure axis be included in the definition of the figure axis in the present study. 
That is, we define the figure axis by the intrinsic figure axis but not by the effective figure axis.
This definition of the figure axis is essentially not different from the generally accepted one.

Anyway, if we know the location of the figure axis thus defined, that of the effective figure axis can be found easily
by following the procedure inferred by the description in the next subsection.   

\subsection{Equations describing the rotational axis relative to the figure axis}

Now we consider the relation between the rotational axis and the figure axis defined in the previous subsection.
Before discussing the subject, however, it will be significant to glance over the nature of the polar motion in the rigid Earth as well as in the deformable Earth
which the rigid dynamics describes.

As mentioned in Section 1, in the free rotation of the rigid Earth, the point on the Earth's surface at which the rotational axis pierces the surface rotates around
the figure axis with the angular speed $\displaystyle \frac{C - A}{A} \omega$,
$\omega$ being the sidereal angular speed of the Earth's rotation (Kinoshita, 1977).
This circular motion of the rotational axis on the surface of the rigid Earth is nothing else but the Euler motion.

% This relation also holds instantaneously at any moment for the deformable Earth as if the deformed Earth were rigid
% if we regard the effective figure axis as the figure axis of the rigid Earth equivalent to the deformed Earth
% and the observed values of $A$ and $C$ (which are a result of the deformation) as those of the equivalent rigid Earth (Kubo, 2009).
A similar circular motion of the rotational axis occurs also in the deformable Earth.
Suppose that the Earth is deformed by the centrifugal force due to the rotation and then consider a rigid Earth with the same values of $A$ and $C$ as in the deformed Earth  
and its figure axis coincident with the effective figure axis, i.e., a rigid Earth equivalent to the deformed Earth.
Then the rotational axis of the deformed Earth behaves in the same way as in the equivalent rigid Earth at each instant (Kubo, 2009).

\begin{figure}
\begin{center}
 \includegraphics[width = 5cm, clip]{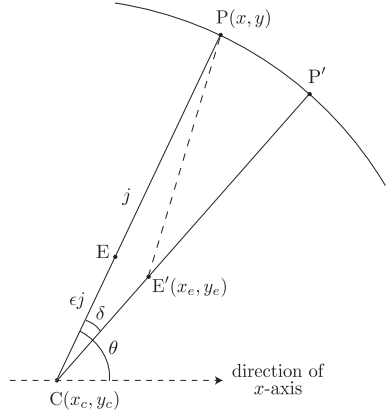}
\end{center}
\caption{The Earth's surface near the north pole.  C and E are the intrinsic and the effective figure axes at some instant, respectively.
P is the rotational axes at the same moment and P$'$ is that some time duration before.
The origin of xy-coordinate system is at IRP (International Reference Pole) and x- and y-axes in the directions of Greenwich and $90^{\circ}$ east longitude, respectively.}
\label{NPole1}
\end{figure}%

To return to the subject, we consider first the Earth which has elasticity but not plasticity.
Fig. \ref{NPole1} shows the Earth's surface near the north pole. Let C and P be the figure axis and the rotational axis at some instant
and let $\overline{\textrm{C} \textrm{P}} = j$.
The bulge by the centrifugal force due to the rotation is produced at the equatorial zone of the rotational axis.  Acoording to Kubo (1991), the figure axis of the Earth
produced as the conpound of the intrinsic Earth's figure and the bulge due to the centifugal force, i.e., the effective figure axis, is located
at the point E on the line CP and at the distance $\epsilon j$ from C with
\begin{equation}
\epsilon = \frac{3\mu C}{C - A},
% \label{beta}
\end{equation}
where $\mu$ is given by
\begin{equation}
\mu = \frac{ka_e^2 {\omega}^2}{9GC} = 0.00116k,
\end{equation}
$k$ being the Love number and $a_e$ the equatorial radius of the Earth, respectively, and $G$ the universal gravitational constant.

Then, as mentioned above concerning the behavior of the rotational axis toward the effective figure axis in the deformable Earth,
P moves so as to rotate around E with the angular speed
\begin{equation}
\Omega = \frac{C - A}{A}\omega. 
\end{equation}
It should be noticed that $A$ and $C$ in this equation is those for the actual Earth shape including the centrifugal effect,
that is, the observed values of the moments of inertia.  Therefore the value of $\Omega $ is $(1/304)\omega $, or the corresponding rotational period is
about 304 sidereal days.  This is the rotational speed of the Euler motion in the rotation of the equivalent rigid Earth.

As a consequence that P makes sush a motion, P rotates around C with an angular speed
\begin{equation}
\Omega ' = (1 - \epsilon)\Omega = \frac{(1-3\mu)C - A}{A}\omega. 
\end{equation}
Supposing $k = 0.29$, this angular speed corresponds to the period of about 440 days, the period of the Chandler wobble.
Thus the period of the Euler motion is prolonged to that of the Chandler wobble due to the elasticity of the Earth,
being coincident with the expression in Kubo (1991) which was derived by a more analytical method.

Next we consider the Earth which has anelasticity besides elasticity.
We treat this condition by introducing a model such that
the bulge by the centrifugal force at a moment is produced around P$'$, the rotational axis some time interval ago, but not around P, the rotational axis at the moment.
Let $\angle \textrm{P}\textrm{C}\textrm{P}' = \delta $.  Then the effective figure axis at the moment comes to the point E$'$ on the line CP$'$ and
at a distance $\epsilon j$ from C. 
We consider that $j$, and therefore $\epsilon j$ as well, does not change during the time interval $\delta /\omega $
corresponding to the motion from P$'$ to P, which is probably one day or so.
This results in that P moves so as to draw a circle around E$'$ with the radius $\overline{\textrm{E}'\textrm{P}}$,
which is equal to $(1 - \epsilon)j$ with a sufficient approximation, and with the angular speed $\Omega $.

This situation is expressed by the following equations:
We consider a coordinate system on the Earth's surface in Fig. \ref{NPole1}.  Although this coordinate system is arbitrary, it may be convenient to choose
International Reference Pole (IRP), which can be regarded as practically the same as Conventional International Origin (CIO), as the origin
and to take x-axis in the longitude of Greenwich.  Then y-axis is in $90^{\circ}$ east longitude.

Let the coordinates of P, C and E$'$ be $(x, y), (x_c, y_c)$ and $(x_e, y_e)$, respectively.  Then the velocity of P relative to E$'$ is written as
\begin{equation}
\begin{split}
\dot{x} &= -(y - y_e)\Omega, \\
\dot{y} &= (x - x_e)\Omega .
\end{split}
\end{equation}
We then express $(x_e, y_e)$ in terms of $(x_c, y_c)$ in these equations.  Let the angle of the direction of the line CP from x-axis be $\theta$.  Then 
\begin{equation}
\begin{split}
x_e &= x_c + \epsilon j \cos(\theta - \delta)\\
    &= x_c + \epsilon j \left( \frac{x - x_c}{j} \cos \delta + \frac{y-y_c}{j}\sin \delta \right) \\
    &\cong x_c + \epsilon[(x- x_c) + (y - y_c)\delta].
\end{split}
\end{equation}
Therefore we have with a sufficient accuracy
\begin{equation}
\begin{split}
x - x_e &= x - x_c - \epsilon[(x - x_c) + (y - y_c)\delta] \\
      &= (1 - \epsilon)(x - x_c) - \epsilon(y - y_c)\delta.
\end{split}
\end{equation}
Similarly,
\begin{equation}
y - y_e = (1 - \epsilon)(y - y_c) + \epsilon(x - x_c)\delta.
\end{equation}
Hence,
\begin{equation}
\begin{split}
\dot{x} &= -[(1 - \epsilon) (y - y_c) + \epsilon (x - x_c)\delta]\Omega, \\ 
\dot{y} &= [(1 - \epsilon) (x - x_c) - \epsilon (y - y_c)\delta]\Omega. 
\end{split}
\label{basic_eq}
\end{equation}
These are the equations for the polar motion which gives the motion of the rotational axis P$(x, y)$
when the location of the figure axis $\textrm{C}(x_c, y_c)$ is given as a known function of time
in the Earth with the elasticity and anelasticity represented by $\mu$ and $\delta$, respectively,
or inversely gives the position of the figure axis when the polar motion is known.

\section{Artificial polar motion}

Now, as a preliminary study to use the equations derived above, we try to generate a polar motion by means of numerical integration 
giving an arbitrary motion of the figure axis.

\begin{figure}
\begin{center}
 \includegraphics[width = 15cm, clip]{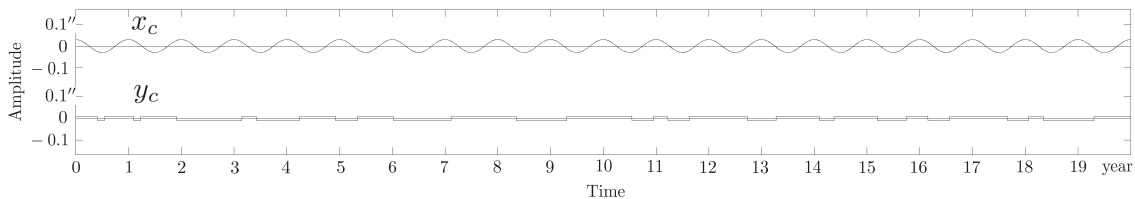}
\end{center}
\caption{The coordinates of the figure axis arbitrarily given for 20 years.
x-coordinate represents the annually periodic nature of the variation of the axis and y-coordinate an irregularity in it.}
\label{fa_given}
\end{figure}

For example, we give a motion of the figure axis as follows:
\begin{equation}
\begin{split}
x_c &= 0.03'' \cos \nu t, \\
y_c &= 0.01'' \times f(t),
\end{split}
\label{afa}
\end{equation}
where $\nu$ is the angular speed corresponding to the period of one tropical year, i.e., $2\pi/365.2422$ days.
$f(t)$ is a step function which takes the values of either $+1$ or $-1$ changing them with random durations of time of $50 \times i~\textrm{days}~(i = 1~\textrm{to}~9)$.
The motion of the figure axis (\ref{afa}) is shown in Fig. \ref{fa_given}.

The variations in $x_c$ is to give an annual variation in the generated polar motion as seen in the actual polar motion
while $f(t)$ in $y_c$ is to give an irregularity to the generated polar motion.
Without this function, the generated polar motion would be a simple periodic motion when $\delta = 0$
and would converge to a periodic motion even if $\delta \ne 0$\footnote{
If $x_c$ and $y_c$ take the form of $x_c = a \sin \nu t + b \cos \nu t$ and $y_c = c \sin \nu t + d \cos  \nu t$ in Eq. (\ref{basic_eq}),
the equations have the following general solution:
\begin{equation*}
\begin{split}
x &= C_1 e^{-\epsilon \delta \Omega t} \cos [(1 - \epsilon)\Omega t + C_2] + p \sin \nu t + q \cos \nu t,\\
y &= C_1 e^{-\epsilon \delta \Omega t} \sin [(1 - \epsilon)\Omega t + C_2] + r \sin \nu t + s \cos \nu t,
\end{split}
\end{equation*}
where $C_1$ and $C_2$ are the constants of integration
and $p, q, r$ and $s$ are linear combinations of $a, b, c$ and $d$ with the coefficients being functions of $\epsilon, \delta, \Omega$ and $\nu$.
If $\delta = 0$, obviously $x$ and $y$ are periodic functions.
If $\delta \ne 0$ (and $\delta>0$), the exponential terms in the solution attenuate and $x$ and $y$ converge to periodic functions, too.
}.
The annual term and the irregular term in the motion of the figure axis are separated into $x_c$ and $y_c$ in order to make it easy
to conferm how faithfully the figure axis is reproduced from the polar motion in the next section. 

As for the initial coordinates of the rotational axis P, we set them as
\begin{equation}
%   \begin{split}
x = +0.1'', ~~~~~y = 0.0''. 
% \end{split}
\end{equation}
These initial values are arbitrary but not given so that the generated polar motion fits the actual one in the Earth, that is, the generated polar motion
is quite fictitious.

We also give the Love number 0.29 which corresponds to the values of $\mu = 0.000336$ and $\epsilon = 0.308$ in performing numerical integration.
The corresponding period of Chandler wobble is about 440days.
As for the value of $\delta$ which is a parameter related to the damping coefficient of the Earth, we first adopt it as 0.

\begin{figure}
\begin{center}
 \includegraphics[width = 15cm, clip]{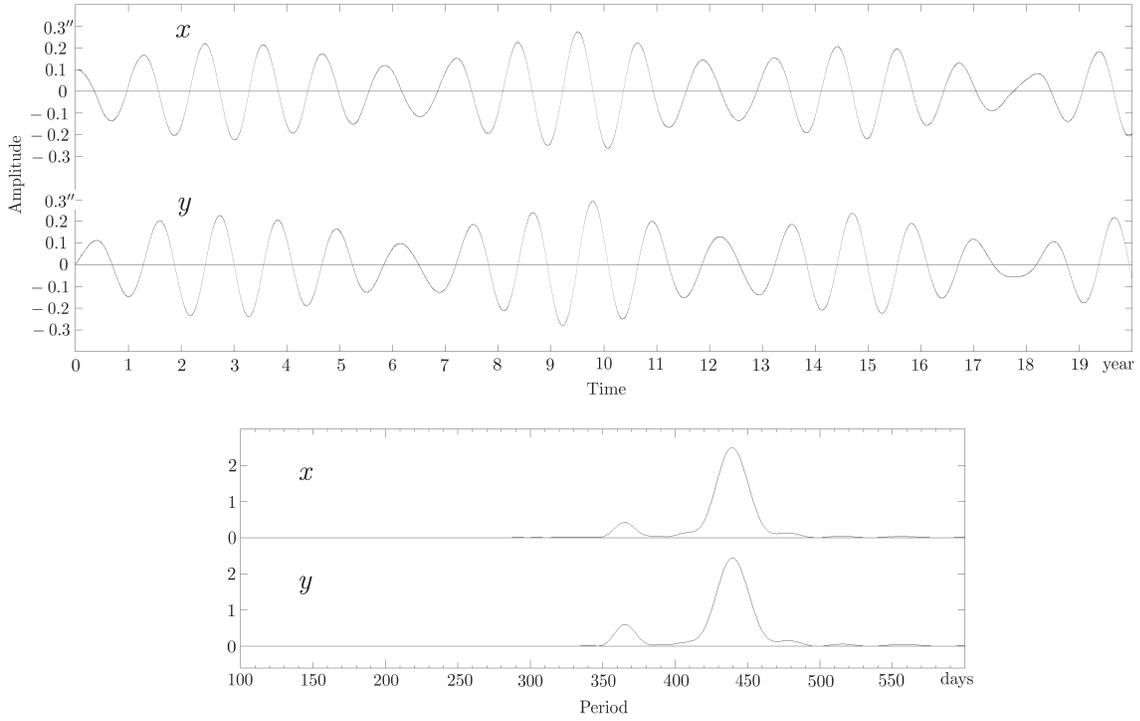}
\end{center}
\caption{The time variation of the coordinates and its power spectra
for an artificially produced polar motion with the figure axis given in Fig. \ref{fa_given} and assuming $\delta = 0$.}
\label{apm0}
\end{figure}%

\begin{figure}
\begin{center}
 \includegraphics[width = 15cm, clip]{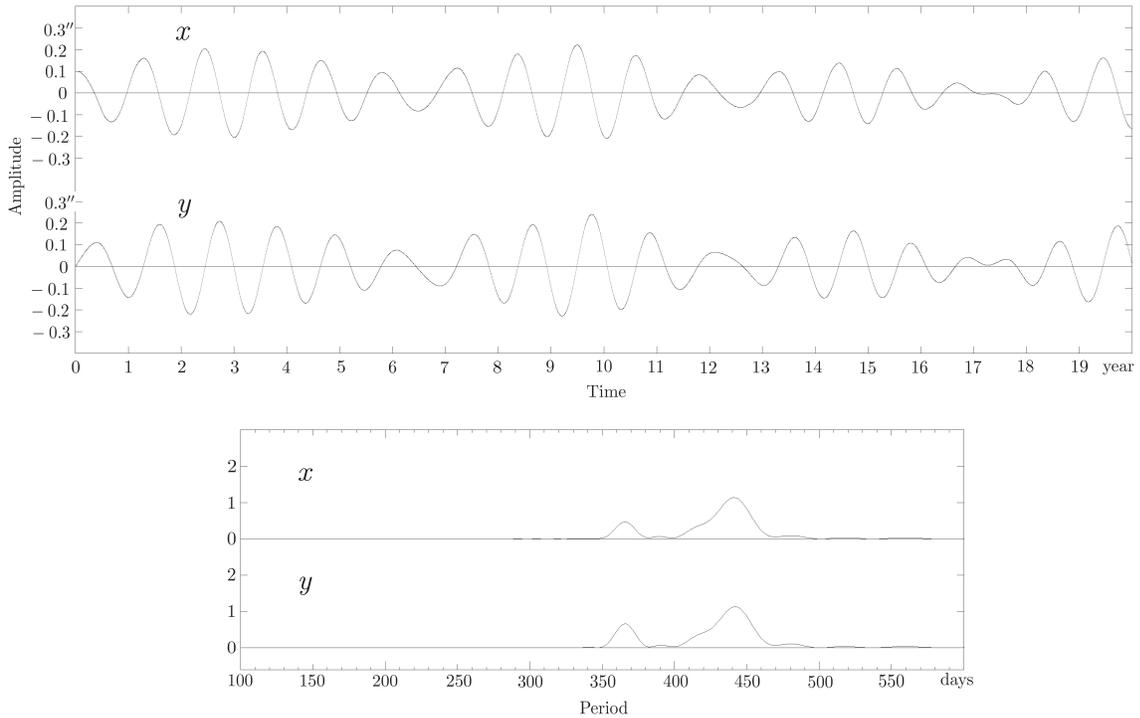}
\end{center}
\caption{The time variation of the coordinates and its power spectra
for an artificially produced polar motion with the figure axis given in Fig. \ref{fa_given} and assuming $\delta = 1^{\circ}$.}
\label{apm1}
\end{figure}%

A numerical integration by the classical Runge-Kutta method for the time span of 20 years (7305 days)
gives a polar motion, i.e., a variation of the coordinates of the rotational axis $(x,y)$ as shown in Fig. \ref{apm0}.
The power spectra for the variation are also shown in the figure.

We next carry out the same numerical integration but changing the value of $\delta$ to $1.0^{\circ}$.
The variation of $(x,y)$ and their spectra in this case are shown in Fig. \ref{apm1}.

From the result of the numerical integration above we see a difference between the polar motions obtained assuming $\delta = 0$ and $\delta = 1.0^{\circ}$,
but it is difficult to judge which represents the actual polar motion better from this result.
Due to the irregular function in (\ref{afa}), the free oscillation does not attenuate in both cases.
In other words, it would be difficult to determine the value of $\delta$ from the data of the actual polar motion.

We further notice in the spectra both in Figs. \ref{apm0} and \ref{apm1}
that the period of the free oscillation appearing in the generated polar motion is close to 440 days, which is the same as the assumed value in the numerical integration.
This fact is not necessarily self-evident.  Although it may be possible to prove it theoretically anyhow, we can say that we have shown it empirically here.

\section{To reproduce the figure axis from the generated polar motion}

Next we consider an inverse problem of the one in the previous section.
That is, contrary to the process to generate a polar motion giving a motion of the figure axis,
we consider the problem to obtain the location of the figure axis at each instant from the produced polar motion.

\begin{figure}
\begin{center}
 \includegraphics[width = 7cm, clip]{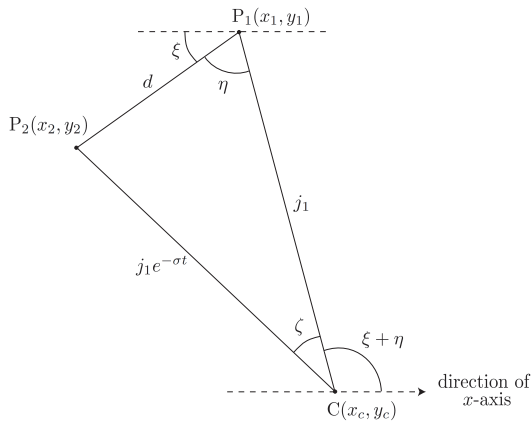}
\end{center}
\caption{The same area as in Fig. \ref{NPole1}.  ${\textrm{P}}_1$ and ${\textrm{P}}_2$ are the rotational axes
at time $t_1$ and $t_2$ ($t_1 < t_2)$.  The figure axis C is supposed to be constant during the time duration $t_1$ to $t_2$.}
\label{NPole2}
\end{figure}

Fig. \ref{NPole2} shows the same area as Fig. \ref{NPole1}.  Let $\textrm{C}(x_c, y_c)$ represent the figure axis at an instant $t_1$
and $\textrm{P}_1(x_1, y_1)$ and $\textrm{P}_2(x_2, y_2)$ the rotational axes at  $t_1$ and $t_2$, $t_2$ being the instant $t$ days later than $t_1$.
We approximate that the figure axis does not move during the period of $t$ days and the rotational axis draws a quasi circle around the figure axis
decreasing the radius gradually due to the attenuation\footnote{
If $\overline{\textrm{C}\textrm{P}_1} = j$, then $\overline{\textrm{C}\textrm{P}_2} = j e^{-\sigma t}$, $\sigma$ being the damping coefficient defined by $-\dot{j}/j$.
According to Kubo (1991), $\sigma = 1.12 \times 10^{-4}\delta/$day, $\delta$ being expressed in degree of angle.
$\delta = 1^{\circ}$ corresponds to the damping with the half-life of about 24.5 years.  For $t = 10$ days, for example,
$\overline{\textrm{C}\textrm{P}_2} = 0.9989 \times \overline{\textrm{C}\textrm{P}_1} $. 
}. 

Define angles $\zeta, \eta$ and $\xi$ as in Fig \ref{NPole2}.
Assume $T$ (days) which is the period of the circular motion of the rotational axis P around the figure axis C
corresponding to the angular speed of P at the instant, and let $\overline{\textrm{P}_1\textrm{P}_2}$ be $d$.  Then $\zeta$ is approximately expressed by  
\begin{equation}
\zeta =\frac{2\pi d}{T} ~~(\textrm{radian}).
\label{zeta}
\end{equation}
Then, from the cosine formula for the plane triangle $\textrm{C} \textrm{P}_1 \textrm{P}_2$,
\begin{equation}
d^2 = j_1^2 + j_1^2 e^{-2\sigma t} - 2j_1^2 e^{-\sigma t} \cos \zeta,
\end{equation}
$j_1$ being $\overline{\textrm{C}\textrm{P}_1}$.  From this 
\begin{equation}
j_1 = \frac{d}{\sqrt{1 + e^{-2\sigma t} - 2 e^{-\sigma t} \cos \zeta}}.
\end{equation}
Applying the sine formula to the same triangle: 
\begin{equation}
\frac{\sin \eta}{j_1 e^{-\sigma t}} = \frac{\sin \zeta}{d},
\end{equation}
we have $\eta$.  Meanwhile $\xi$ is obtaied by
\begin{equation}
\tan \xi = \frac{y_1 - y_2}{x_1 - x_2}.
\end{equation}
Usind $\eta$ and $\xi$ thus obtained, we have the coodinates of C by 
\begin{equation}
\begin{split}
x_c &= x_1 - j_1 \cos (\xi + \eta),\\
y_c &= y_1 - j_1 \sin(\xi + \eta).
\end{split}
\end{equation}
These coordinates are regarded as those of the figure axis at the time $\displaystyle \frac{t_1 + t_2}{2}$.

Now, following this procedure, we obtain the locations of the figure axis for every day from the data of the artificial poler motion generated in the previous section.
In doing this, of the two polar motions we obtained with different values of $\delta$, we adopt the one given in Fig. \ref{apm1},
which is generated with the value of $\delta = 1.0^{\circ}$.

Further, in order to perform this, it is necessary to assume the period of the free oscillation $T$ in Eq. (\ref{zeta}).
Redgarding this, we may assume 440 days which has been obtained in the spectrum analysis (Fig. \ref{apm1}).
As seen in the previous section, the value appearing in the spectrum shows a very good coincidence with the value given for the period of the free oscillation
when producing the artificial polar motion.

As for the damping coefficient $\delta$ (therefore for $\sigma$), on the other hand, we can not know the true value for the actual Earth.
So, concerning the artificial polar motion, we analyze it first assuming  the value of 0 and then assuming the value $1^{\circ}$,
the latter being the same as the one used in the simulation, i.e., the true value so to speak.  

\begin{figure}
\begin{center}
 \includegraphics[width = 15cm, clip]{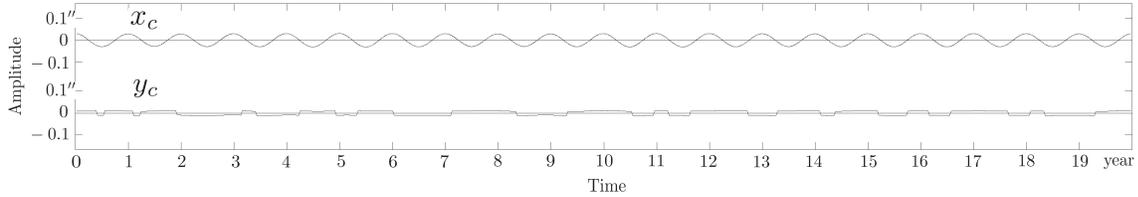}
\end{center}
\caption{The coordinates of the reproduced fgure axis from the polar motion shown in Fig. \ref{apm1} and assuming $\delta = 0$.}
\label{fa_got_1_0}
\end{figure}%

\begin{figure}
\begin{center}
 \includegraphics[width = 15cm, clip]{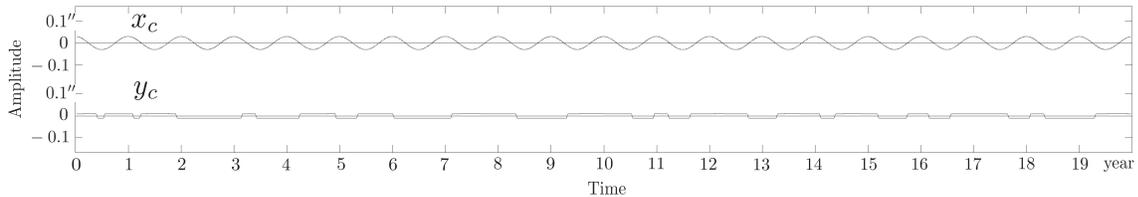}
\end{center}
\caption{The coordinates of the reproduced fgure axis from the polar motion shown in Fig. \ref{apm1} and assuming $\delta = 1^{\circ}$.}
\label{fa_got_1_1}
\end{figure}%

The result is shown in Figs. \ref{fa_got_1_0} and \ref{fa_got_1_1}, for $\delta = 0$ and $\delta = 1^{\circ}$, respectively.
In the two figures, Figs. \ref{fa_got_1_0} and \ref{fa_got_1_1}, we do not see a large difference between the cases assuming $\delta = 0$ and $\delta = 1.0^{\circ}$,
and the original motion of the figure axis given by Eq. (\ref{fa_given}) is reproduced very well in both cases,
although we admit a slight superiority of the reproduction in the case where the true value is assumed for the damping parameter.
This suggests that the choice for the value of the damping parameter which must be assumed in the process of obtaining the figure axis from the data of the polar motion
does not affect the final result so much.

Anyway, it is shown through this example that the figure axis can be determined from the polar motion with a very good accuracy.

\section{The polar motion in the actual Earth and its primitive analysis}

In the following sections we treat the actual polar motion observed in the Earth.  The data of the polar motion or the coordinates of the rotational axis is acquired
as Bulletin B data published by the International Earth Rotation Service (IERS), which can be downloaded from the internet website (IERS, 2012).
Of these data we use the daily data for the period of 20 years (7305 days) from 1 January 1992 to 31 December 2011.
It is emphasized that they are all compiled based on the modern observation data acquired using the space-geodetic techniques.

\begin{figure}
\begin{center}
 \includegraphics[width = 15cm, clip]{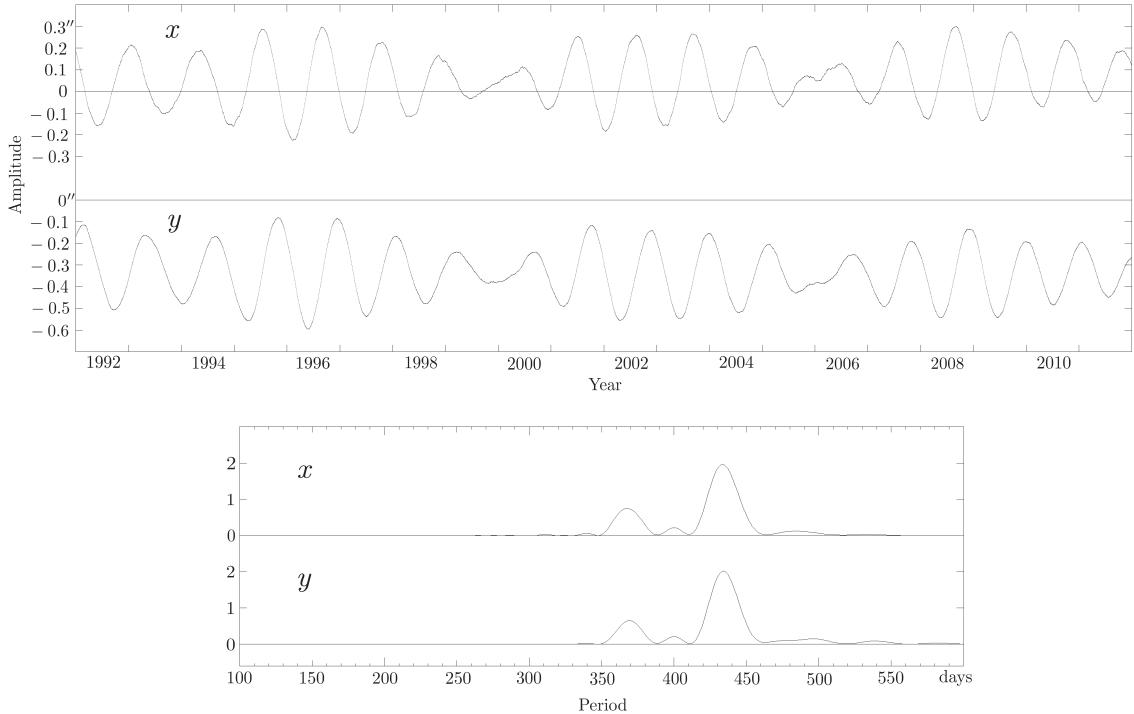}
\end{center}
\caption{The time variation of the coordinates and its power spectra for the actual polar motion observed in the Earth for the recent 20 years.}
\label{PM}
\end{figure}

First we look at the original data of the coordinates of the rotational axis $(x, y)$, which is shown in Fig. \ref{PM}.  The origin of the coordinate system is
the International Reference Pole (IRP) and x- and y-axes are taken in the directions of the Greenwich longitude and $90^{\circ}$ east longitude.
So, attention should be paid to that the sign of y-coordinate is opposite to that in the IERS original data.
The power spectra of x- and y-coordinates are shown also in Fig. \ref{PM}

The polar motion consists of a secular motion and periodic motions.
Concerning the periodic variation, they consist of mostly the free oscillation (the Chandler wobble), the annual and the semi-annual components,
among which the semi-annual component is comparatively small.

Looking at the spectra, we notice first the prevailing Chandler period of about 437 days.
And, other than this main component, the annual and semi-annual components are recognized.  On the whole all these features are consistent with many preceding analyses,
although some minor differences may exist due to the difference of the period for the analysis and so on.

Regarding the coordinates $(x, y)$ of the IERS data shown in Fig. \ref{PM}, we express them as functions of time assuming the quadratic polynomial
for the secular motion and fitting respective ellipses for the periodic motions of the Chandler wobble and the annual and semi-annual terms.
That is, we perform the least squares fitting assuming the expressions for $x$ and $y$, respectively, as
\begin{equation}
\begin{split} 
x &= a_0 + a_1 t + a_2 t^2\\
  &+ c_1 \sin \Omega't + c_2 \cos \Omega't + c_3 \sin \nu t + c_4 \cos \nu t + c_5 \sin 2 \nu t + c_6 \cos 2 \nu t, \\
y &= b_0 + b_1 t + b_2 t^2\\
  &+ d_1 \sin \Omega't + d_2 \cos \Omega't + d_3 \sin \nu t + d_4 \cos \nu t + d_5 \sin 2 \nu t + d_6 \cos 2 \nu t,
\end{split}
\label{func}
\end{equation}
where $t$ is the time beginning from 1 January 1992, expressed in one tropical year, i.e., 365.2422 days.
Also $\Omega' = (2\pi \times 365.2422 / 437)$ radian/year and $\nu = 2\pi$ radian/year.

\begin{figure}
\begin{center}
 \includegraphics[width = 8.5cm, clip]{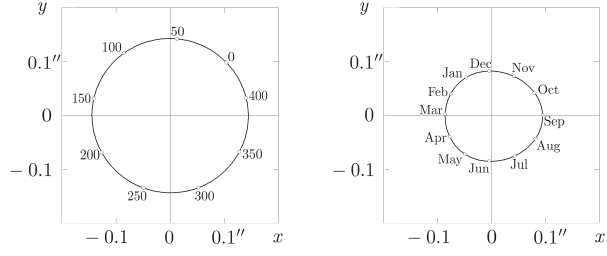}
\end{center}
\caption{The Chandler (left) and the annual (right) components in the polar motion averaged for 20 years, the semi-annual component being incorporated in the latter.
The 0-th day in the Chandler component corresponds to 1 January 1992 and the days after every Chandler period.
The names of the months in the annual component represent the first day of each month.}
\label{PMComp_Ch}
\end{figure}%

The result is: \\
For the secular motion,
\begin{equation}
\begin{array}{ll}
a_0 = [+0.034 (\pm 0.002)]'', &b_0 = [-0.318 (\pm 0.002)]'', \\
a_1 = [+0.0022 (\pm0.0008 )]''/t, &b_1 = [-0.0023 (\pm 0.0008)]''/t,\\
a_2 = [-0.00036 (\pm0.00009 )]''/t^2, &b_2 = [-0.00013 (\pm 0.00009)]''/t^2.
\end{array}
\label{poly_pm}
\end{equation}
For the Chandler wobble with the period of 437 days,
\begin{equation}
\begin{array}{ll}
c_1 = [-0.1001 (\pm 0.0007)]'', &d_1 = [+0.1036 (\pm 0.0007)]'', \\
c_2 = [+0.1039 (\pm 0.0007)]'', &d_2 = [+0.0986 (\pm 0.0007)]''. \\
\end{array}
\label{Ch_pm}
\end{equation}
For the annual variation,
\begin{equation}
\begin{array}{ll}
c_3 = [-0.0777 (\pm 0.0007)]'', &d_3=  [-0.0424 (\pm 0.0007)]'', \\
c_4 = [-0.0462 (\pm 0.0007)]'', &d_4 = [+0.0722 (\pm 0.0007)]''. \\
\end{array}
\label{an_pm}
\end{equation}
For the semi-annual variation,
\begin{equation}
\begin{array}{ll}
c_5 = [+0.0043 (\pm 0.0007)]'', &d_5 = [+0.0005 (\pm 0.0007)]'', \\
c_6 = [-0.0009 (\pm 0.0007)]'', &d_6 = [-0.0014 (\pm 0.0007)]''. \\
\end{array}
\label{s-an_pm}
\end{equation}

Eqs. (\ref{Ch_pm}) to (\ref{s-an_pm}) give the locii of the free oscillation and the annual and the semi-annual terms and they are shown in Fig. \ref{PMComp_Ch},
in which the semi-annual component is included in the annual one.
The two locii exhibit nearly circular patterns.

Here, notice should be taken that the patterns are the averaged ones for the duration of 20 years from 1992 to 2011.
That is, the periodic components are assumed roughly constant throughout the period.
% All these results both for the secular motion and the periodic variation are essentially the same as those in the preceding works.
% Since both the phenomenon and the process of analysis are simple, it is not likely that the results are quite different.
As for the secular motion we consider the problem below again.

\section{The position and the motion of the figure axis obtained from the polar motion}  

Now we apply the procedure described in Section 4 to the actual polar motion shown in Fig. \ref{PM}
and obtain the positions of the figure axis on the Earth's surface for every day during 20 years.
As for the period of the free oscillation in the polar motion we assume 437 days, the same as obtained in the spectra in Fig. \ref{PM},
since we know that the period of the free oscillation to assume in Eq. (\ref{zeta}) appears as the peak in the spectra as we saw in Section 4.

Also we do not know the value for $\delta$.  As for it we assume 0.
Concerning this we saw also in Section 4 that the error in this choice does not affect the result so much.

\begin{figure}
\begin{center}
 \includegraphics[width = 15cm, clip]{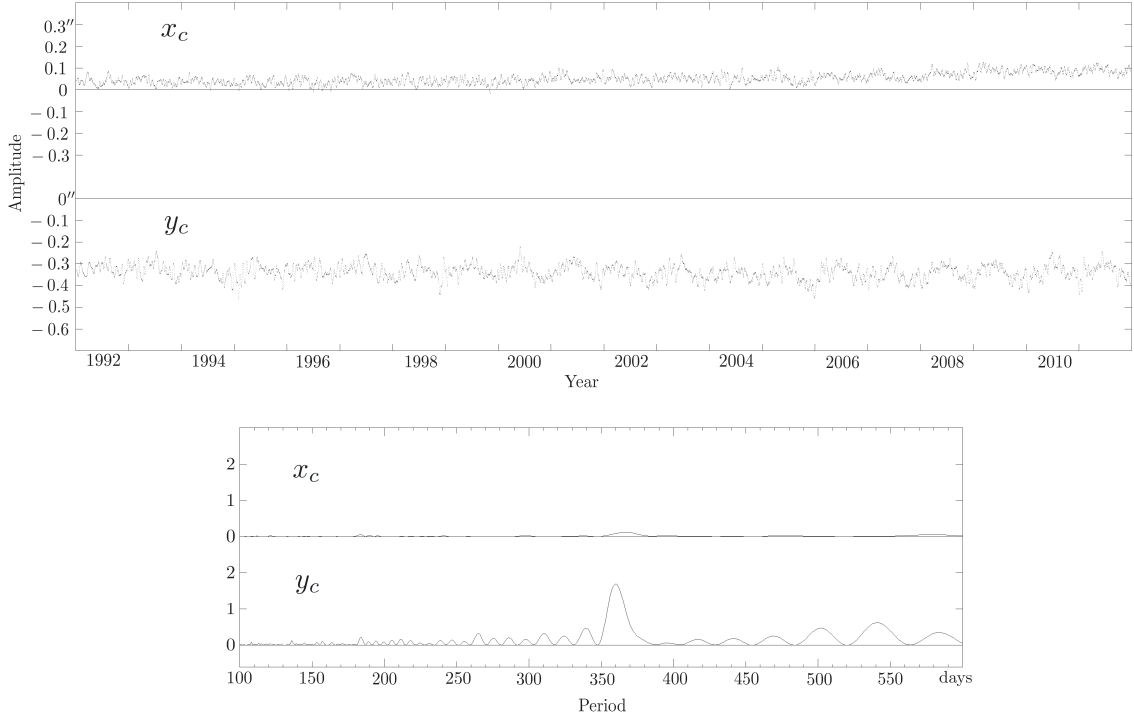}
\end{center}
\caption{The time variation of the coordinates and its power spectra for the figure axis obtained from the polar motion shown in Fig. \ref{PM} and assuming $\delta = 0$.}
\label{FA}
\end{figure}%

The obtained values of the coordinates $x_c$ and $y_c$ for every day and the spectra
for the periods contained in their variations are shown in Fig. \ref{FA}.
Remarkably, there exists no component with the period of the free oscillation around 437 days at all in the spectra.
This should be natural because it is not likely that the variation of the figure axis is affected by the free oscillation.
Therefore, in finding the expressions as functions of time for those coordinates by carrying out the least squares fitting,
we assume forms similar to Eq. (\ref{func}) but not containing the oscillation with the period of 437 days, i.e., not containing $c_1, c_2, d_1$ and $d_2$:
\begin{equation}
\begin{split} 
x_c &= a_0 + a_1 t + a_2 t^2\\
  &+ c_3 \sin \nu t + c_4 \cos \nu t + c_5 \sin 2 \nu t + c_6 \cos 2 \nu t, \\
y_c &= b_0 + b_1 t + b_2 t^2\\
  &+ d_3 \sin \nu t + d_4 \cos \nu t + d_5 \sin 2 \nu t + d_6 \cos 2 \nu t,
\end{split}
\end{equation}

The result is: \\
For the secular motion,
\begin{equation}
\begin{array}{ll}
a_0 = [+0.035 (\pm 0.001)]'', &b_0 = [-0.327 (\pm 0.001)]'', \\
a_1 = [+0.0004 (\pm0.0003 )]''/t, &b_1 = [+0.0011 (\pm 0.0006)]''/t,\\
a_2 = [-0.00001 (\pm0.00004 )]''/t^2, &b_2 = [-0.00045 (\pm 0.00007)]''/t^2.
\end{array}
\label{poly_faxis}
\end{equation}
For the annual variation,
\begin{equation}
\begin{array}{ll}
c_3 = [+0.0085 (\pm 0.0003)]'', &d_3=  [+0.0102 (\pm 0.0005)]'', \\
c_4 = [+0.0004 (\pm 0.0003)]'', &d_4 = [-0.0236 (\pm 0.0005)]''. \\
\end{array}
\label{an_faxis}
\end{equation}
For the semi-annual variation,
\begin{equation}
\begin{array}{ll}
c_5 = [-0.0001 (\pm 0.0003)]'', &d_5 = [+0.0038 (\pm 0.0005)]'', \\
c_6 = [-0.0046 (\pm 0.0003)]'', &d_6 = [+0.0065 (\pm 0.0005)]''. \\
\end{array}
\label{s-an_faxis}
\end{equation}

In examining the result in the analysis in this section in reference to that in the previous section,
we first recognize an innegligible difference of about $0.01''$ between the constant terms in $(x,y)$ and $(x_c, y_c)$ given by
Eqs. (\ref{poly_pm}) and (\ref{poly_faxis}), respectively.
Concerning that, we must say that the averaged position of $(x_c, y_c)$ has a significant physical meaning but that of $(x, y)$ does not necessarily.
As for the secular motions, on the other hand, it will be difficult to say something significant.

\begin{figure}
\begin{center}
 \includegraphics[width = 4cm, clip]{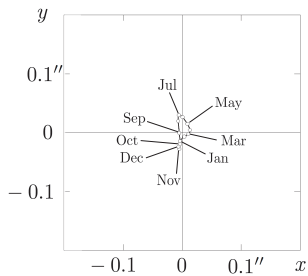}
\end{center}
\caption{The locus for the motion of the figure axis in the Earth, averaged for 20 years.
The names of the months represent the first day of each month.}
\label{FAComp_ann}
\end{figure}

Next, when we compare the annual and the semi-annual variations in $(x,y)$ and $(x_c, y_c)$, we see a striking difference.
The amplitudes of the annual and semi-annual components in $(x_c, y_c)$ given by Eqs. (\ref{an_faxis}) and (\ref{s-an_faxis}) are much smaller
than those in $(x, y)$ given by Eqs. (\ref{an_pm}) and (\ref{s-an_pm}). 
Further, the locus of the annual and the semi-annual components given by Eqs. (\ref{an_faxis}) and (\ref{s-an_faxis})
and shown in Fig. \ref{FAComp_ann} 
is not like a circle but rather close to a line compared to that in the polar motion.

Here again it should be noticed that the locus is the averaged one for 20 year duration. 
The shape changes from year to year and, other than that, there exist many small variations with shorter periods.

Since the modern observation of the polar motion has an accuracy as high as $0.0001''$ and the time resolution as short as one day,
the position of the figure axis obtained from the polar motion data is considered to have a corresponding good accuracy.
Although only the variation of the figure axis in a global time scale has been investigated here, the most of the details in the variation
should not be attributed to the errors in the observations or in the process of the analysis but should be considered to reflect the dynamical state of the Earth.
Therefore they are necessary to be examined more fully in future studies.

\section{A model to explain the variation of the figure axis}

In this section we propose a model of the Earth which brings about the variation of the figure axis obtained in the previous section and shown in Fig. \ref{FAComp_ann}.
In doing this, it would be a touchstone to judge the validity of the model if it produces a reasonable magnitude as well as 
a rather linear locus regarding the variation of the axis.

We now consider a case where a seasonal transfer of matter in the vertical direction occurs in some region on the Earth's surface. 
Most of the matter may be water and the transfer may take place either in plants or in the air, or in both.

Suppose that the total mass to move is $m$ and the transfer occurs in some area around a point with the longitude and latitude $(\lambda, \phi)$,
and let the vertical distance of the transfer $h(t)$ from the mean height be expressed by
\begin{equation}
 h(t) = l \sin(\nu t + {\psi }),
\label{h}
\end{equation}
$\psi$ being an arbitrary phase angle.

With respect to a geocentric rectangular coordinate system with x-, y-axes in the directions of the Greenwich longitude and $90^{\circ}$ east longitude
on the equator, respectively, the coordinates of the point are
\begin{equation}
x = a \cos \phi \cos \lambda,~~~~~ y = a \cos \phi \sin \lambda,~~~~~z = a \sin \phi,
\end{equation}
and the components of the transfer vector of the matter, $\Delta x, \Delta y$ and $\Delta z$, are given by
\begin{equation}
\Delta x = h(t) \cos \phi \cos \lambda ,~~~~~\Delta y = h(t) \cos \phi \sin \lambda,~~~~~\Delta z = h(t) \sin \phi,
\end{equation}
where $a$ is the Earth's radius at the point but it may be regarded as equal to the equatorial radius $a_e$.

If the moments of inertia for x-, y- and z-axes be respectively $A, B$ and $C$, then the changes of them due to this transfer of the matter are
\begin{equation}
\begin{split}
\Delta A &= \Delta m(y^2 + z^2) = 2m(y\Delta y + z \Delta z) = 2mah(t)(1-{\cos}^2 \phi{\cos}^2 \lambda), \\
\Delta B &= \Delta m(z^2 + x^2) = 2m(z\Delta z + x \Delta x) = 2mah(t)(1-{\cos}^2 \phi{\sin}^2 \lambda), \\
\Delta C &= \Delta m(x^2 + y^2) = 2m(x\Delta x + y \Delta y) = 2mah(t){\cos}^2 \phi.
\end{split}
\end{equation}
At the same time, the products of inertia for x-, y- and z-axes, respectively denoted by $D, E$ and $F$,
which are originally 0, are generated as follows;
\begin{equation}
\begin{split}
D &= -\Delta (myz) = -m(y\Delta z + z\Delta y) = -2mah(t)\sin\phi\cos\phi\sin \lambda, \\
E &= -\Delta (mzx) = -m(z\Delta x + x\Delta z) = -2mah(t)\sin\phi\cos\phi \cos \lambda, \\
F &= -\Delta (mxy) = -m(x\Delta y + y\Delta x) = -2mah(t){\cos}^2\phi \sin \lambda \cos \lambda.
\end{split}
\end{equation}

We next investigate how the location of the figure axis in the Earth shifts accompanying to this change of the moments and the products of inertia.
We notice that this change of the location of the figure axis is similar to that caused by the deformation in the elastic Earth due to the body tide
originating from the attraction of the outer bodies, studied by Kubo (2009) in detail.  So we follow this preceding study in the following.

In the non-deformed Earth or the avarage shape of ther Earth, let the axes which are associated with the moments of inertia $A, B$ and $C$ be denoted by A-, B- and C-axes,
respectively, and those of the deformed Earth by A$'$-, B$'$- and C$'$-axes, respectively.

\begin{figure}
\begin{center}
 \includegraphics[width = 8cm, clip]{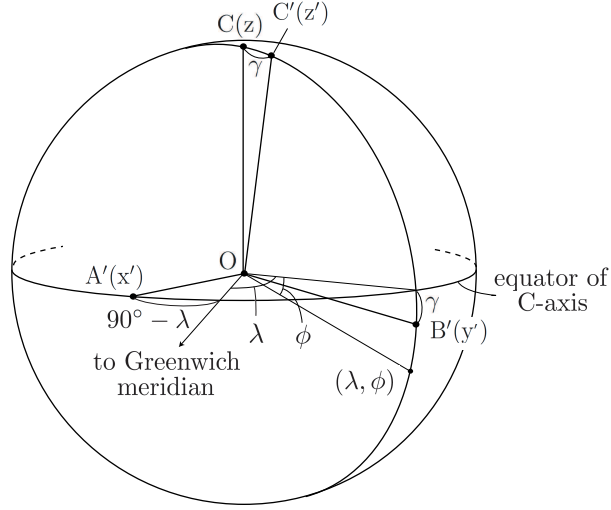}
\end{center}
\caption{The averaged (ABC) and the deformed (A$'$B$'$C$'$) figures of the Earth.  The figure is represented by A-, B- and C-axes which are associated with
the moments of inertia, $A, B$ and $C$, respectively.  $\gamma$ is the shift of C-axis due to the deformation.
It is so assumed that the vertical transfer of natter around the point $(\lambda, \phi)$ causes the deformation.}
\label{fgamma}
\end{figure}

The shifted new figure axis, i.e., C$'$-axis, is expected to be located on the meridian plane of the longitude $\lambda$,
and one of the other new principal axes, B$'$-axis, is also on this plane and separated from C$'$-axis by $90^{\circ}$. 
% Let the right ascension of the body in the celestial equatorial coordinate system be $\alpha$, which is measured from the equinox along the celestial equator.
% Then $\alpha  = \hat {\alpha} + g + l + 180^{\circ}$, $\hat {\alpha}$ being the longitude in the body fixed coordinate system as above.
% Both the right ascensions of C$'$- and B$'$-axes are $\alpha$,
Then the other new principal axis, A$'$-axis, is on the original equator and in the longitude $\lambda - 90^{\circ}$ (Fig. \ref{fgamma}). 

Consider a new coordinate system with x$'$-, y$'$- and z$'$-axes coinciding with A$'$-, B$'$- and C$'$-axes, respectively.
% As a result of this deformation, the actual principal axes of inertia in the Earth are produced at different places from the averaged principal axes.  
% Thus induced C$'$-axis (z$'$-axis) and another induced principal axis (B$'$-axis or y$'$-axis) are expected to be located in the plane
% containing the original C-axis (z-axis) and the outer body.

% Let the right ascension of the body in the celestial equatorial coordinate system be $\alpha$, which is measured from the equinox along the celestial equator.
% Then $\alpha  = \hat {\alpha} + g + l + 180^{\circ}$, $\hat {\alpha}$ being the longitude in the body fixed coordinate system as above.
% Both the right ascensions of C$'$- and B$'$-axes are $\alpha$,
% and the induced A$'$-axis (x$'$-axis) is in the xy-plane and at the direction of right ascension $\alpha - 90^\circ$ (Fig. \ref{fgamma}). 
Let the co-latitude of z$'$-axis in xyz-system be $\gamma $. 
The values of the moments and the products of inertia with respect to x$'$y$'$z$'$ coordinate system, $A', B', C', D', E'$ and $F'$,
are obtained by examining how the equation for the quadratic surface (the inertia ellipsoid):
\begin{equation}
(A+\Delta A)x^2 +(A+\Delta B)y^2 +(C + \Delta C)z^2 + 2Dyz + 2Ezx + 2Fxy = 1
\end{equation}
is changed to the new equation of the quadratic surface: 
\begin{equation}
A'x'^2 + B' y'^2 + C' z'^2 + 2D'y'z' + 2E'z'x' + 2F'x'y' = 1,
\end{equation}
after the rotation of the coordinates sytem from xyz to x$'$y$'$z$'$ given by 
\begin{equation}
\left(\begin {array} {c} x \\ y \\ z \end {array} \right)
 = R_z(90^\circ - \lambda)R_x(\gamma ) \left( \begin {array} {c} x' \\ y' \\ z' \end {array} \right),
\label{rot}
\end{equation}
where $R_i(\xi )$ means the rotation around $i$-axis by the angle $\xi $.

Explicitly (\ref{rot}) is written as
\begin{equation}
\begin{split}
\left(\begin {array} {c} x \\ y \\ z \end {array} \right)
 &= \left( \begin{array}{c}\sin \lambda x' + \cos \lambda \cos\gamma y'+\cos \lambda \sin\gamma z' \\
-\cos \lambda x' + \sin \lambda \cos\gamma y'+\sin \lambda \sin\gamma z'\\
-\sin\gamma y' + \cos\gamma z' \end{array} \right) \\
 &= \left( \begin{array}{c}\sin \lambda x' + \cos \lambda y'+ \gamma \cos\lambda z'\\
-\cos \lambda x' + \sin \lambda y'+ \gamma \sin \lambda z'\\
-\gamma y' + z' \end{array} \right) + O(\gamma ^2).
\end{split} 
\end{equation}

Then, taking into account that $\gamma$ is the same order as $mah(t)/C$, we have after some manipulation
\begin{equation}
A' = A + 2mah(t) + O({\gamma}^2),
\end{equation} 
\begin{equation}
B' = A + 2mah(t) {\sin}^2 \phi + O({\gamma}^2),
\end{equation} 
\begin{equation}
C' = C + 2mah(t){\cos}^2 \phi + O({\gamma}^2),
\end{equation} 
\begin{equation}
D' = 2\gamma (A - C) - 4mah(t)\sin \phi \cos \phi + O({\gamma}^2),
\label{getgamma}
\end{equation} 
\begin{equation}
E' = O({\gamma}^2),
\end{equation} 
\begin{equation}
F' = O({\gamma}^2).
\end{equation} 

In (\ref{getgamma}), if $\gamma $ satisfies the following relation:
\begin{equation}
\gamma = \frac{2mah(t)}{A - C}\sin \phi \cos \phi ,
\label{gamma}
\end{equation}
then $D' = O({\gamma}^2)$ and therefore x$'$-, y$'$- and z$'$-axes are the new principal axes of inertia to the accuracy of $O(\gamma)$.
% Concerning the magnitudes of the moments of inertia $A, B (= A)$ and $C$, they change only by $O(c/C)$.
% Since $c/C \cong 10^{-8}$ (Kubo, 1991), we can regard as $A' = A, ~B' = B ~(= A)$ and $C' = C$ in some cases in the following.

According to the discussion above, we can conclude that the position of the figure axis shifts with respect to its averaged position by the angular distance $\gamma$
and in the direction of the longitude $\lambda$. 

We suppose, e.g., that $\displaystyle \frac{ml}{m_e a} = 2 \times 10^{-10}$, where $m_e$ is the mass of the Earth, and $\phi = -15^{\circ}$. 
Then, taking into account $C \cong 0.33 m_e a_e^2$, we have  
\begin{equation}
\gamma \cong 0.02'' \sin(\nu t + \psi),
\end{equation}
almost the same magnitude of the variation in the figure axis we obtained in the previous section.

If we assume $l = 6 \textrm{m}$, supposing the case of the mass transfer in plants, then $\displaystyle \frac{m}{m_e} \cong 2 \times 10^{-4}$ 
and if $l = 6000 \textrm{m}$, supposing the transfer in the air, then $\displaystyle \frac{m}{m_e} \cong 2 \times10^{-7}$.
% At any rate, it is concluded that a considerably large amount of mass moves.
 
Of course, the transfer of the mass does not occur only at one place but at many places.
In this sense $m$ may be written as $\sum m$ symbolically.
And actually, it would be occurring in all the places on the Earth's surface and there would be yielded an averaged transfer which does not depend on the longitude, too.
Then each local transfer which brings about the variation of the figure axis could be regarded as a deviation from the averaged transfer. 
% The global transfer independent of the longitude dose not cause the variation of the figure axis.
% Only local transfers generate the changes of the figure axis and the transfer at one plase brings about a linear change of the position of the figure axis.

The actual variation of the figure axis is the vector sum of many variations each caused by a local deviation from the averaged transfer.
The shape of such a sum would be a line if all the local transfers are synchronous, i.e., if $\psi$ in Eq. (\ref{h}) is equal at all points,
but generally be an ellipse with considerable eccentricity. 

% Next we see what variation in the rotational speed is brought about by the global transfer of the mass distribution with the same model. 

On the other hand, a global vertical transfer of the mass independent of longitude causes a change of $C$ and consequently a change of the Earth's ratational speed.
For simplicity, consider a case where such a global transfer occurs uniformly in the zone along the equator with the mass per unit area $m'$ and the total mass $M$,
and the amount of the transfer follows the same form as Eq. (\ref{h}).
Then, very roughly, $\Delta C(t)$, the variation of $C$ due to this transfer of mass, is given by
\begin{equation}
\Delta C = \int 2m'ah(t) d\phi d\lambda = 2ah(t)\int m'd\phi d\lambda = 2Mah(t).
\end{equation}
The consequent variation of the rotational speed $\Delta \omega$ is
\begin{equation}
\Delta \omega = \Delta \frac{H}{C} = -\frac{H}{C^2}\Delta C = -\frac{\Delta C}{C}\omega = -\frac{2Mah(t)}{C}\omega = -\frac{2Mal'}{C} \omega \sin (\nu t + \psi'),
\label{omega1}
\end{equation}
where $H$ is the total angular momentum of the rotation which is constant and $\omega$ is the rotational speed.

On the other hand, the seasonal variation of the rotational angle, or that of the universal time, is known to be comparatively stable
and the following expression has been used for a considerablly long time (USNO and HMNAO, 2000):
\begin{equation}
\Delta T_s = 22 \sin \nu t - 12 \cos \nu t - 6 \sin 2\nu t + 7 \cos 2\nu t~~(\textrm{ms}). 
\label{Ts}
\end{equation}

Differentiating this equation we have for the variation of the rotational speed only with regard to the annual terms,
\begin{equation}
\begin{split}
\Delta \omega  &= (4.4 \times 10^{-9} \cos \nu t + 2.4 \times 10^{-9} \sin \nu t) \omega + \dots \\
               &= 5.0\times 10^{-9}  \omega \sin (\nu t + 61^{\circ})+ \dots.
\end{split}
\end{equation}

Comparing this with Eq. (\ref{omega1}),
we see that they coincide roughly when we assume $\displaystyle \frac{Ml'}{m_e a} = 8 \times 10^{-10}$ and $\psi ' = 241^{\circ}$,
where we have taken into acount $C \cong 0.33m_e a_2^2$ again.
Thus, assuming $l' = l$, the ratio of the masses involved in the local deviations of the transfer to that in the global transfer is  
\begin{equation}
\frac{\sum m}{M} \cong 0.25.
\end{equation}

This value of the ratio is reasonable.
Therefore it is natural to conclude that the variation of $\Delta T_s$ is caused by the globally averaged transfer of mass
while the motion of the figure axis by the local deviations of the transfer.

\section{Conclusions}

The variation of the figure axis reflects the physical state in the Earth more directly than the polar motion
since the change inside the Earth first brings about a variation of the figure axis and then it affects the polar motion.

In the present study, the equatons to give the relation between the figure axis and the rotational axis in the Earth
with both elasticity and anelasticity have been derived.
Since the polar motion is the motion of the rotational axis on the Earth's surface, the obtained relation makes it possible to determine the
position of the figure axis in the Earth-fixed frame at any moment from the data of the polar motion.

% Preliminarily, using the equations, an artificial polar motion has been generated and its characteristics is examined for the Earth with the same properties
% of elasticity and anelasticity.
%
% Both cases where the damping coefficient exists and does not have been examined,
% but any clear difference has not been recognized between both cases.
% As far as an irregularity is contained in the variation of the figure axis, no damping of the free oscillation is seen in both cases.

The position and the motion of the figure axis obtained from the actual data of the polar motion exhibits various characteristics.

First, the averaged position of the polar motion are not coincident with that of the figure axis exactly
but is dislocated  by about $0.01''$, while the differnce between the secular motions of both is not so definite.

The periodic variation of the figure axis has obvious and comparatively stable components of annual and semi-annual periods 
but no component with the period of the free oscillation.

On the whole, the annual and the semi-annual components draw a shape resembling a thin ellipse with the semi-major axis of about $0.02''$.

While the figure axis draws a locus like this as the average for a long interval, the shape changes from year to year and,
other than that, there exist many minor variations with shorter periods.
Those minor variations are considered to be significant but not due to the errors in the observations or in the process of the analysis,
and therefore they are necessary to be studied more in detail.

Finally, a simple model with the seasonal flow of the matter in the Earth has been introduced to explain the variation of the figure axis obtained above. 
Applying the same model also to the variation of the rotational speed, it is confirmed that the variation estimated is accordant with the one which is so far known,
thus suggesting that the variations of the figure axis and of the rotational speed are brought about by the same cause as far as the seasonal variation is concerned.


\begin{thebibliography}{}

\bibitem{dickman}  Dickman, S. R. 1993.  Dynamic ocean-tide effects on Earth's rotation.  Geophys. J. Int., \textbf{112}, 448 - 470
                   DOI 10.1111/j.1365-246X.1993.tb01180.x

\bibitem{Gibert}   Gibert, D. and Le Mou$\ddot{\textrm{e}}$l, J.-L. 2008. Inversion of polar motion data: Chandler wobble, phase jumps, and geomagnetic jerks.
                   J. Geophys. Res., \textbf{113}, B10405, DOI 10.1029/2008JB005700

\bibitem{gross}    Gross, R. S. 2000. The excitation of the Chandler wobble.
                   Geophys. Res. Lett., \textbf{27}, 2329 - 2332
        
\bibitem{guinot}   Guinot, B. 1972. The Chandlerian wobble from 1900 to 1970.
                   Astron. Astrophys., \textbf{19}, 207 - 214
   
\bibitem{hophner}  H$\ddot{\textrm{o}}$pfner, J. 2002. Chandler and annual wobles based on space-geodetic measurements.
                   Sci. Tech. Rep. Geoforschungzentrum Potsdam, 02/13, 1 - 10
   
\bibitem{IAU}      IAU. 2000. ``Proceedings of the 24th General Assembly Manchester, UK, August 7 - 18, 2000",
                   Transactions of the IAU Vol. XXIV B, Ed. H. Rickman

\bibitem{IERS1}     IERS. 2004. ``Explanatory Supplement to IERS Bulletins A and B",
                   Service de la Rotation Terrestre, Observatoire de Raris and
                   Rapid Service/Prediction Centre, U.S. Naval Observatory.

\bibitem{IERS2}     IERS. 2012. Standard EOP data files, finals.data (IAU2000).\\ 
                   http://www.iers.org/IERS/EN/DataProducts/EarthOrientationData/eop.html

\bibitem{kimura}    Kimura, H. 1917. Variations in the fourteen months' component of the polar motion.
                    Monthly notices R. Astron. Soc., \textbf{78}, 163 - 167
  
\bibitem{kinoshita} Kinoshita, H. 1977. Theory of the rotation of the rigid Earth.
                    Celest. Mech. Dyn. Astron. \textbf{15}, 277 - 326

\bibitem{kubo1}     Kubo, Y. 1991. Solution to the rotation of the elastic Earth by method of rigid dynamics.
                    Celest. Mech. Dyn. Astron. \textbf{50}, 165 - 187

\bibitem{kubo2}     Kubo, Y. 2009. Rotation of the elastic Earth.  The role of the angular-velocity-dependence
                    of the elasticity caused perturbation.
                    Celest. Mech. Dyn. Astron. \textbf{105}, 261 - 274, DOI 10.1007/s10569-009-9225-2

\bibitem{munk}       Munk, W.H. and MacDonald, G.F.J. 1960.  ``The Rotation of the Earth", Cambridge Univ. Press, London 

\bibitem{newcomb}    Newcomb, S. 1892. On the dynamics of the Earth's rotation, with respect to the periodic variations of latitude.
                     Manthly Notice R. Astron. Soc., \textbf{52}, 336 - 341
   
\bibitem{seidelmann} Seidelmann, P.K. 1982. 1980 IAU theory of nutation: the
                     final report of the IAU working group on nutation.
                     Celest. Mech. Dyn. Astron. \textbf{27}, 79 - 106

\bibitem{USNO}     USNO and HMNAO. 2000. ``Explanatory Supplement to the Astronomical Almanac",
                   Nautical Almanac Office, U. S. Naval Obs. and H. M. Nautical Almanac Office, R. Greenwich Obs.,
                   University Science Books, Sausalito, California

% \bibitem{yumi-yokoyama}  Yumi, S. and Yokoyama, K. 1980. Results of the ILS in a homogeneous system 1899.9-1979.0.
%                          Publ. Central Bureau of the International Earth Rotation Service, Internat. Latitude Observ. Mizusawa Japan, 199pp

\end{thebibliography}
\end{document}